\documentclass[twocolumn,showpacs,preprintnumbers,amsmath,amssymb,aps,prl]{revtex4-1} 
\usepackage{dcolumn}
\usepackage{float}
\usepackage{graphicx}
\usepackage{siunitx}
\usepackage{ifpdf}
\usepackage{epstopdf}
 \usepackage[cmyk]{xcolor}
 \usepackage{latexsym}
 \usepackage{mathrsfs}
 \usepackage{titlesec}

\begin{document}
\preprint{APS/123-QED}

\title{Partially Nondestructive Continuous Detection of Individual Traveling Optical Photons}
\author{Mahdi Hosseini, Kristin M. Beck, Yiheng Duan, Wenlan Chen and Vladan Vuleti\'c }

\affiliation{Department of Physics and Research Laboratory of Electronics, Massachusetts Institute of Technology, Cambridge, Massachusetts 02139, USA}

\begin{abstract}
We report the continuous and partially nondestructive measurement of optical photons. For a weak light pulse traveling through a slow-light optical medium (signal), the associated atomic-excitation component is detected by another light beam (probe) with the aid of an optical cavity. We observe strong correlations of $g^{(2)}_{sp}=4.4(5)$ between the transmitted signal and probe photons. The observed (intrinsic) conditional nondestructive quantum efficiency ranges between 13\% and 1\% (65\% and 5\%) for a signal transmission range of 2\% to 35\%, at a typical time resolution of 2.5 $\si{\micro}$s. The maximal observed (intrinsic) device nondestructive quantum efficiency, defined as the product of the conditional nondestructive quantum efficiency and the signal transmission, is 0.5\% (2.4\%). The normalized cross-correlation function violates the Cauchy-Schwarz inequality, confirming the non-classical character of the correlations.
 \end{abstract}

 \maketitle
Photons are unique carriers of quantum information that can be strongly interfaced with atoms for quantum state generation and processing \cite{Dudin:science2012, Peyronel:Nature2012,Lang:prl2011,Fushman:science2008,chen:science2013,Gorniaczyk:PRL2014,Tiarks:prl2014,Michler:science2000, Tanji:PRL2009}. Quantum state detection, a particular type of processing, is at the heart of quantum mechanics and has profound implications for quantum information technologies. Photons are standardly detected by converting a photon's energy into a measurable signal, thereby destroying the photon. Nondestructive photon detection, which is of interest for many quantum optical technologies \cite{Sperling:PRA2014, Nemoto:PRL2004,Imoto:pra1985}, is possible through strong non-linear interactions \cite{Imoto:pra1985} that ideally form a quantum non-demolition (QND) measurement \cite{Grangier:Nature1998}. To date, QND measurement of single microwave photons bound to cooled cavities has been demonstrated with high fidelity using Rydberg atoms \cite{Haroche:Nature1999,Gleyzes:nature2007,Guerlin:Nature2007}, and in a circuit cavity quantum electrodynamics system using a superconducting qubit \cite{Johnson:NatPhys2010}. 
\begin{figure}[!t]
	\centerline{\includegraphics[width=\columnwidth]{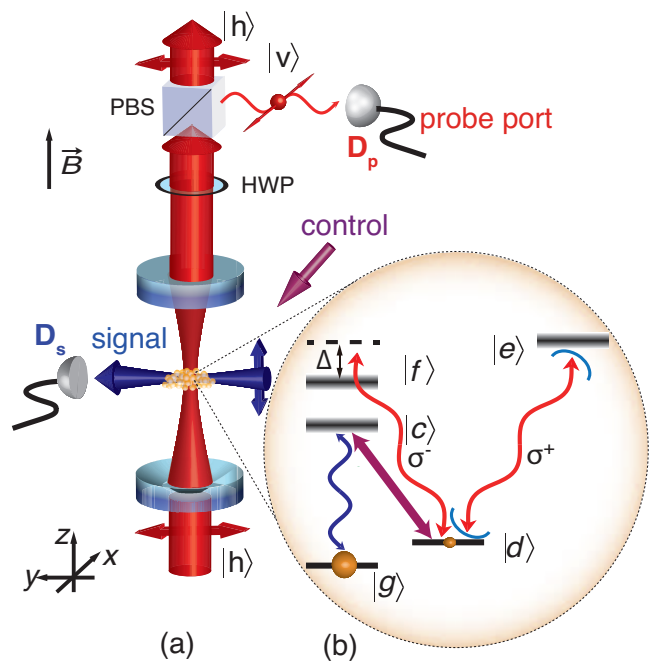}}
	\caption{{\bf (a)}, {\bf (b)}, The $\pi$-polarized signal light travels with slow group velocity through the atoms by means of EIT on the $|g\rangle\leftrightarrow|c\rangle\leftrightarrow|d\rangle$ transitions. The associated atomic-excitation component is nondestructively detected via cavity light in the geometric overlap between the atomic ensemble and the cavity mode. Input cavity light is linearly polarized such that, in the absence of signal photons, the probe port of the polarization beam splitter (PBS) ideally remains dark. Whenever a signal photon traverses the atomic medium in the cavity, the transmission of the $\sigma^+$ polarized light through the cavity is blocked. The atomic levels are $|g\rangle$= $|6S_{1/2}; F=3, m_F=3\rangle$, $|c\rangle$= $|6P_{3/2}; 3, 3\rangle$, $|d\rangle$=$|6S_{1/2}; 4, 4\rangle$, $|e\rangle$= $|6P_{3/2}; 5, 5\rangle$ and $|f\rangle$= $|6P_{3/2}; 5, 3\rangle$, where $F,m_F$ are the hyperfine quantum numbers.}
\label{fig1: setup}
\end{figure}

For quantum communication and many other photonics quantum information applications \cite{Knill:Nature2000, Gisin:Nature2007}, it is desirable to detect traveling optical photons instead of photons bound to cavities. Previously, a single-photon transistor was realized using an atomic ensemble inside a high finesse cavity where one stored photon blocked the transmission of more than one cavity photon and could still be retrieved \cite{chen:science2013}. Such strong cross-modulation \cite{Beck:prl2014} can be used for all-optical destructive detection of the stored optical photon, but the parameters in that experiment did not allow nondestructive detection with any appreciable efficiency. High-efficiency pulsed nondestructive optical detection has recently been achieved using a single atom in a cavity \cite{Rempe:science:2013}. In that implementation, the atomic state is prepared in 250 $\si{\micro}$s, altered by the interaction with an optical pulse reflected from the cavity, and read out in 25 $\si{\micro}$s.

In this Letter, we realize partially nondestructive, continuous detection of traveling optical photons with micro-second time resolution. The signal photons to be detected propagate through an atomic ensemble as slow-light polaritons \cite{Fleischhauer:EIT2000} under conditions of electromagnetically induced transparency (EIT) \cite{Harris:eit97}. The signal polariton's atomic-excitation component is nondestructively detected via the polarization change on another light field (probe), enhanced by an optical cavity. We observe positive correlations between the signal and probe photons of $g_{sp}^{(2)}=4.4(5)$, and use the measured correlation function to calculate the conditional nondestructive quantum efficiency Q. We achieve efficiencies Q between 13\% and 1\% at a signal transmission $T_s$ between 2\% and 35\%, with a maximum device nondestructive quantum efficiency $Q\times T_s$ of 0.47\% at a maximum signal input rate of 300~kHz.
 \begin{figure}[!t]
	\centerline{\includegraphics[width=\columnwidth]{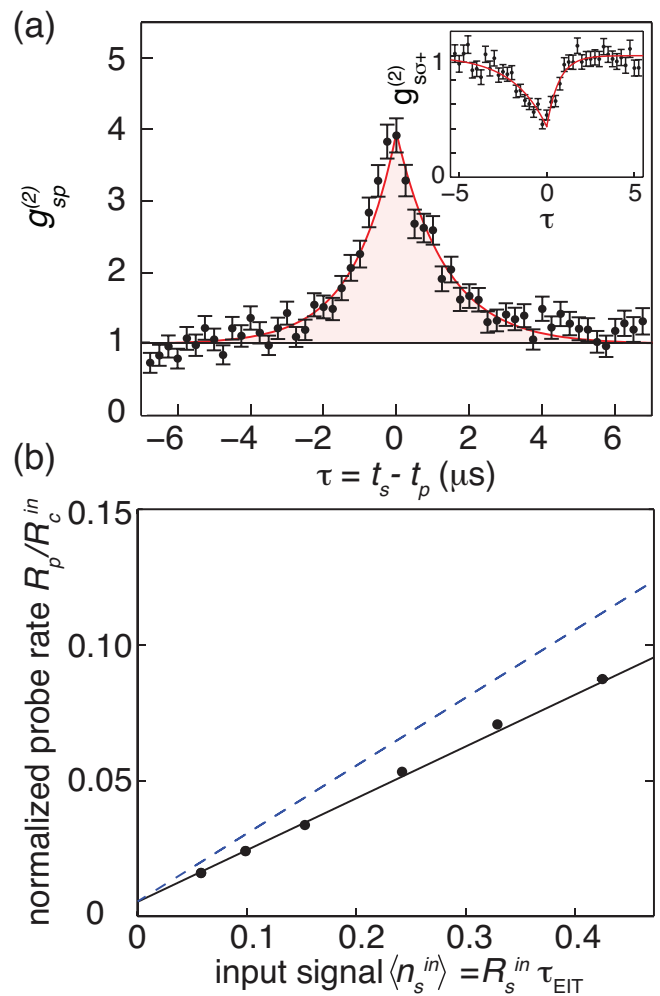}}
	\caption{{\bf (a)} Signal-probe correlation function, $g_{sp}^{(2)}$, is plotted as a function of separation time, $\tau$, between signal ($t_s$) and probe ($t_p$) photons. The decay time constant for negative (positive) times $\tau_{<}=1.2(2)$ $\si{\micro}$s ($\tau_{>}=1.3(2)$ $\si{\micro}$s) is consistent with the cavity decay time (EIT lifetime) \cite{Beck:prl2014}. This measurement is done with mean input cavity photon number $\langle n^{in}_c \rangle=R^{s=0}_c\tau_c/q_c=3.7$, cavity path detection efficiency $q_c=0.2$, $\tau_c=(\kappa/2)^{-1}=2~\si{\micro}$s and Rabi frequency $\Omega/2\pi=1.9$ MHz. The inset shows the cross-correlation function for signal and $\sigma^+$-polarized cavity photons, measured for $\langle n^{in}_c \rangle=0.1$ and $\Omega/2\pi= 2.6$ MHz. The observed signal-probe anti-correlation is $g^{(2)}_{s\sigma^+}(0)=0.41(7)$. In this and all following figures, statistical errorbars are plotted when they are larger than the points and indicate one standard deviation. {\bf (b)} Normalized detected probe rate $R_p/R_c^{s=0}$, plotted against input signal photon number $\langle n^{in}_s \rangle=R^{in}_s\tau_{_{\text{EIT}}}/q_s$. The slope of the solid fitted line is $0.20(1)$. The dashed line represents the maximum possible probe rate with a slope of $\varepsilon_{id}=0.25$. For this measurement, $\langle n^{in}_c \rangle=1.2$, $q_s=0.3$, and $\Omega/2\pi=1.3$~MHz, giving $\tau_{_{\text{EIT}}}=1.4$ $\si{\micro}$s.}
\label{fig2}
\end{figure}
 
The nondestructive measurement scheme and atomic level structure are shown in Fig.~\ref{fig1: setup}. A laser-cooled atomic ensemble of $^{133}$Cs atoms is held in a cigar-shaped dipole trap that partly overlaps with the fundamental mode of the optical cavity.
A signal light resonant with the $|g\rangle\rightarrow|c\rangle$ transition propagates orthogonal to the cavity axis through the ensemble. A control laser induces an EIT transmission window that slows down the signal light to a typical group velocity of $300~m/s$ and reversibly maps it onto a collective atomic excitation in state $|d\rangle$ \cite{Fleischhauer:EIT2000}.  This atomic population couples strongly to the $\sigma^+$ polarized light which is simultaneously resonant with the optical cavity and the $|d\rangle\rightarrow|e\rangle$ transition, blocking its transmission through the cavity \cite{Beck:prl2014,Imamoglu:PRL1997, Soljacic:APL2005}. To generate a useful positive detection signal in transmission, we add $\sigma^-$ reference light and probe the cavity continuously with horizontally polarized light. The reference light interacts only weakly with the atoms: the atomic coupling strength on the $|d\rangle\rightarrow|f\rangle$ transition is 45 times smaller than the strength of the $\sigma^+$ transition and is also detuned from resonance by $\Delta/2\pi=6$~MHz by the $5.2$~G magnetic field along the cavity axis ($z$). Light transmitted through the cavity is then analyzed in a horizontal/vertical basis.  Vertically polarized light (probe port) corresponds to detection, as the probe port is dark in the absence of signal photons.

Quantum correlations between detected outgoing signal and probe photons are the signature of nondestructive detection. The cross-correlation function $g^{(2)}_{sp}=\langle n_sn_p\rangle/\langle n_s\rangle\langle n_p\rangle$ can be understood as the likelihood of measuring the signal twice: first measuring it nondestructively with our cavity QED system, which results in a detected probe photon ($n_p=1$), and then checking the first measurement by measuring the signal photon again destructively ($n_s=1$). The cross-correlation function in  Fig.~\ref{fig2}a with zero-time value $g^{(2)}_{sp}(0)=4.0(3)$ demonstrates that simultaneous nondestructive and destructive measurements of the signal photon occur four times more often than randomly. This value also agrees well with the directly observed blocking of $\sigma^+$-polarized cavity photons by a signal photon (inset to Fig.~\ref{fig2}a), and with the theoretical expectations for our system's cooperativity $\eta=4.3$ and relevant optical depth $\mathcal{D}\simeq3$ (see S.M.).  The increased $\mathcal{D}$ accounts for the improvement over previously published results with the same apparatus \cite{Beck:prl2014}.

To confirm the linearity of the system, we plot the probe rate normalized to the empty cavity output rate, $R_p/R_c^{s=0}$, against the average input signal photon number per EIT lifetime $\langle n_s^{in}\rangle=R_s^{in}\tau_{\text{EIT}}/q_s$ in Fig.~\ref{fig2}b. Here, $R_s^{in}/q_s$ is measured input rate corrected for the finite detection efficiency $q_s=0.3$. Under ideal circumstances, an incident cavity photon emerges in the probe port with probability $\varepsilon_{id}=1/4$ in the presence of a signal photon, indicated in the figure as a dashed line. Achieving this limit requires a strong single-atom-cavity coupling (cooperativity $\eta\gg1$) \cite{Haruka:AAMOP2011}, large ensemble optical depth inside the cavity region $\mathcal{D}\gg1$, and sufficiently slowly traveling signal photons $\tau_{\text{EIT}}/\tau_c>1$, where $\tau_c$ is the cavity lifetime. Even with finite cooperativity $\eta$ and optical depth, we measure $\varepsilon=0.20(1)$. This number is the detection probability per input cavity photon and includes both nondestructive and destructive detection of the signal photon. The nonzero offset in Fig.~\ref{fig2}b at $\langle n_s^{in}\rangle=1$ corresponds to the background noise in the average measurement. The observed linear increase in probe rate for $\langle n_s^{in}\rangle<1$ also confirms the sensitivity of our experiment at the single photon level. However, unlike output correlations, this average signal neither distinguishes between destructive and nondestructive detection events nor does it reveal the time resolution of the detector. Destructive detection events correspond to decohered polaritons, i.e. atomic population in state $|d\rangle$, and hence have the same effect on the cavity light as traveling signal photons. 

To study only those events when we preserve the signal photon, we define the conditional nondestructive quantum efficiency, $\text{Q}$, to be the conditional probability for a correlated photon to be detected in the probe port when a signal photon is present: $\text{Q}=\langle n_sn_p\rangle/\langle n_s\rangle-\langle n_p\rangle$ for $\langle n_s\rangle\ll1$. (Note that the second term $\langle n_s\rangle=\langle n_p\rangle\langle n_s\rangle/\langle n_s\rangle$ is necessary to remove uncorrelated (random) coincidences between signal and probe photons.) The time scale for this conditioning is defined by the typical correlation time: this conditional nondestructive quantum efficiency $\text{Q}$ is precisely the area under $g^{(2)}(\tau)-1$ (the shaded area in Fig.~\ref{fig2}a) multiplied by the average rate of detected photons at the probe port, $R_p$: $\text{Q}= R_p\int{(g^{(2)}(\tau)-1)d\tau}$.  $\text{Q}$ evaluates to 10\% for the cross-correlation function plotted in Fig.~\ref{fig2}a. The time resolution is the sum of the positive- and negative- correlation times,  $\tau_>+\tau_<=2.5(4)~\si{\micro}$s \cite{Beck:prl2014}.  Since  $\text{Q}$ scales with the detected rate at the probe port, finite probe photon detection efficiency $q_p$ directly reduces  $\text{Q}$. The total detection efficiency for probe photons, $q_p=0.2$, is the product of detector efficiency (0.45), fiber coupling and filter losses (0.7) and cavity outcoupling losses (0.66).  Correcting for these linear losses gives the intrinsic conditional nondestructive quantum efficiency $\text{Q}/q_p=50\%$. Single photon detectors with better than 0.99 efficiency exist at our wavelength, so only improving the optics and detectors outside of our vacuum chamber would already allow us to achieve a  $\text{Q}$ of 30\%.
\begin{figure}[!t]
	\centerline{\includegraphics[width=\columnwidth]{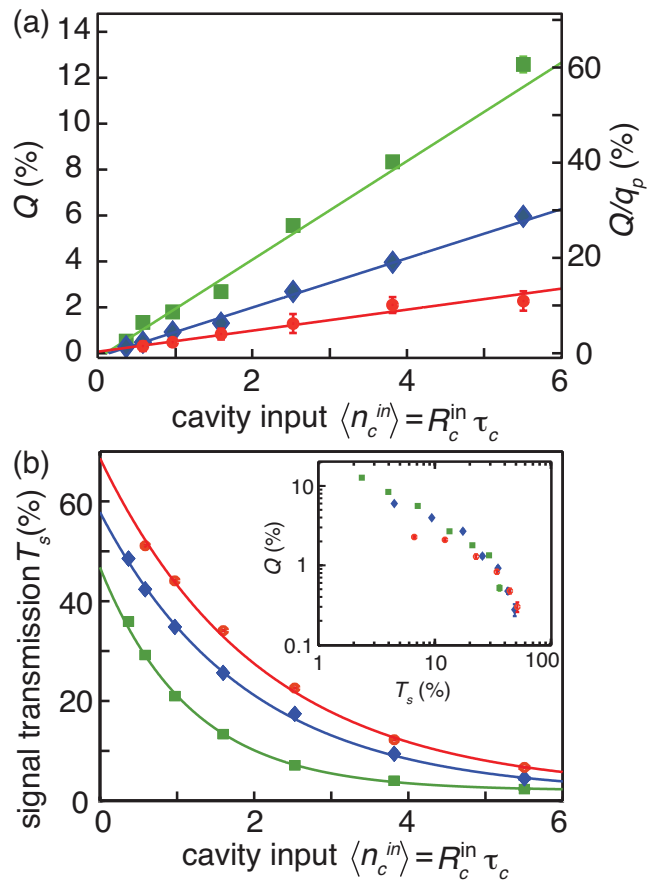}}
	\caption{{\bf (a)} The observed conditional nondestructive quantum efficiency  $\text{Q}$ is plotted against mean cavity photon number, $\langle n^{in}_c\rangle$, with mean $R^{in}_s=2.8\times10^{5}$ s$^{-1}$. The slope of the fitted curves (solid lines) is $\frac{d\text{Q}}{d\langle n^{in}_c\rangle}=\{10(2), 5(1), 1.9(5)\}$\% for $\Omega/2\pi= \{1.8, 2.9, 3.5\}$ MHz (top to bottom) and represents the observed detection efficiency per input cavity photon. {\bf (b)} Signal transmission $T_s$ for the same data presented in (a). Exponential fits give $1/e$ transmission at cavity photon numbers of $\{1.2(1),1.9(1), 2.1(1)\}$ for $\Omega/2\pi= \{1.8, 2.9, 3.5\}$ MHz (bottom to top), respectively. The inset displays the nondestructive quantum efficiency Q as a function of signal transmission for the same Rabi frequencies as in (a) and (b).}
\label{fig3}
\end{figure}
We define the device nondestructive quantum efficiency as the probability for an input photon to be nondestructively detected. It is equal to  $\text{Q}\times T_s$, the product of the conditional nondestructive quantum efficiency and the signal transmission. Fig.~\ref{fig3} explores the tradeoff between these two factors.  $\text{Q}$ scales linearly with the input cavity photon number (Fig.~\ref{fig3}a), as with increasing cavity input rate it becomes more likely for a randomly arriving cavity photon to ``hit" a signal photon and perform the detection. At the same time, the signal transmission, $T_s$, degrades exponentially with input cavity rate due to cavity-induced decoherence of the signal polariton, as seen in Fig.~\ref{fig3}b. Slower signal polaritons (smaller control Rabi frequency $\Omega$) are more likely to be ``hit" by a cavity photon, and thus have a larger nondestructive quantum efficiency but also have a lower transmission due to greater decoherence for a given cavity photon number. The choice of Rabi frequency changes the detector speed but does not improve the tradeoff between efficiency and transmission; the inset of Fig.~\ref{fig3}b shows that observed quantum efficiency as a function of signal transmission collapses to a single curve for all measured Rabi frequencies.

\begin{figure}[!t]
	\centerline{\includegraphics[width=\columnwidth]{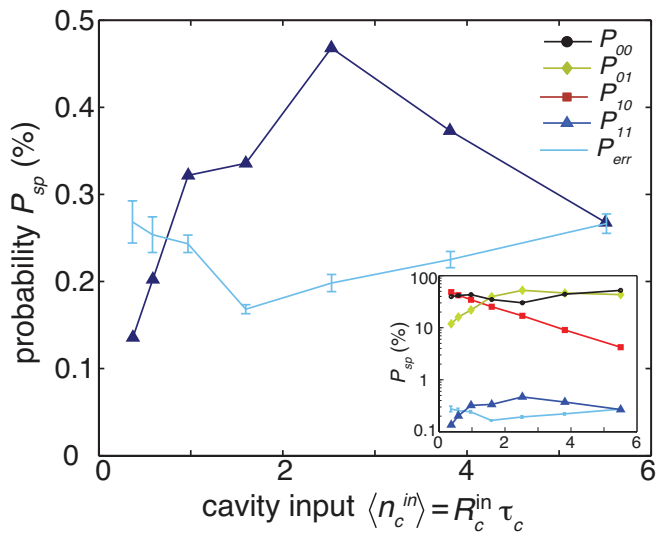}}
	\caption{Device nondestructive quantum efficiency ($P_{11}=\text{Q}\times T_s$) and error probability $P_{\text{err}}$ for joint detection of signal and probe outputs for $\Omega/2\pi=2.9$ MHz. $P_{11}$ is calculated assuming one input signal photon. $P_{\text{err}}$ represents the false detection of probe photons in absence of signal photons. Inset displays all four characterizing probabilities $P_{sp}$ with $s,p=\{0,1\}$ (see text) and $P_{\text{err}}$ under the same conditions. }
\label{fig4}
\end{figure}

Fig.~\ref{fig4} plots the device nondestructive quantum efficiency and the error probability, $P_{err}$ as a function of input  photon number. The maximal observed (intrinsic) device nondestructive quantum efficiency is 0.47\% (2.4\%). $P_{err}$ is the probability of having a false detection event when no signal photon is present. Considering these together, the detector achieves a nondestructive signal to noise ratio of 2.4. We further characterize the performance for single input photons by calculating by the four probabilities, $P_{sp}$, to detect $s=\{0,1\}$ signal and $p=\{0,1\}$ probe photons given one input signal photon. These probabilities can be obtained from measured quantities in the limit $\langle n^{in}_s\rangle\ll1$ using the relations $P_{11}/(P_{11}+P_{10})=\text{Q}$, $P_{11}+P_{10}=T_s$, $P_{01}+P_{11}=\langle n_p\rangle/\langle n^{in}_s\rangle$ and $\sum{P_{sp}}=1$. These probabilities describe different aspects of the nondestructive detection. In particular, the device nondestructive quantum efficiency is $P_{11}$ and the state preparation probability, $P_{11}/(P_{11}+P_{01})\simeq4\%$, represents the probability of having an outgoing signal photon if a photon is present at the probe port. 

In our system, the transmission of detected signal photons is limited to about 70\% by the standing wave nature of our cavity probe, which imprints a grating onto the detected polariton and reduces its readout efficiency in the original mode. In addition, since the atomic medium extends outside the cavity mode, the detection localizes the signal polariton in a finite region of the ensemble, and the corresponding spectral broadening outside the EIT transmission window reduces the signal transmission by 30\% (see S.M.). Finally, cavity photon scattering into free-space, which destroys the signal polariton, occurs with a finite probability $2\eta/(1+\eta)^2=0.3$. The combination of these effects explains the observed transmission reduction for the signal. 

To further enhance the effective optical density of the medium, we also carried out an experiment where the signal propagates twice through the medium (see S.M.). In this case, we observed slightly stronger correlations of $g^{(2)}_{sp}(\tau=0)=4.4(5)$ due to the larger effective optical depth. To remove classical correlations from the observed cross-correlation, we normalize the cross-correlation function to the auto-correlations measured at the signal and probe ports of $g_{ss}^{(2)}=1.6(3)$ and $g_{pp}^{(2)}=5.6(1)$. The resulting normalized quantum correlation $G_{sp}=\left(g_{sp}^{(2)}\right)^2\left/\left(g_{ss}^{(2)}g_{pp}^{(2)}\right)\right.=2.7(8)$ at $\tau=0$ violates the Cauchy-Schwarz inequality \cite{Clauser:PRD1974}, $G<1$, and confirms that our interactions are non-classical. 

Key to the nondestructive photon measurement scheme demonstrated here is the strong interaction between one atom and a cavity photon \cite{McKeever:nature2003, Volz:nature2011,Shomroni:science2014} (large single-atom cooperativity $\eta\gg1$), in combination with the strong collective interaction of atoms with signal photons (large optical depth $\mathcal{D}\gg1$ inside the cavity). Both quantities can be further improved in our experiment. For realistic values $\mathcal{D}=10$ and $\eta=20$, we expect a device nondestructive quantum efficiency exceeding 55\% with a conditional nondestructive quantum efficiency of about 80\% and a signal transmission of about 70\%. An interaction of this kind enables many quantum applications such as the projection of a coherent state of a light pulse into a photon number state \cite{Guerlin:Nature2007}, the implementation of nearly deterministic photonic quantum gates through nondestructive measurement and conditional phase shift \cite{Nemoto:PRL2004}, engineering exotic quantum states of light \cite{Wang:pra2005} or non-deterministic noiseless amplification for entanglement distillation \cite{Zhang:PRA2012}. 

The authors would like to thank Arno Rauschenbeutel for insightful discussions. This work was supported by NSF and the Air Force Office of Scientific Research. K.M.B. acknowledges support from NSF IGERT under grant 0801525.

\bibliographystyle{plainbib}
\bibliography{Ref}

\clearpage
\section*{Supplemental Material}

\makeatletter \renewcommand{\thefigure}{S\@arabic\c@figure} \renewcommand{\thetable}{S\@arabic\c@table}  \makeatother
\makeatletter
\setcounter{figure}{0}
\renewcommand*{\@biblabel}[1]{[S#1]}
\renewcommand*{\@cite}[1]{[S#1]}
\makeatother
\normalsize

{\bf {Methods}}
Each second-long experimental cycle has a 12~ms detection period, which consists of 20~$\si{\micro}$s measurement times, a time window arbitrarily chosen to be much longer than the EIT lifetime to allow the continuous measurement of signal photons, interleaved with 20~$\si{\micro}$s preparation times that ensure the atoms are optically pumped to the $|g\rangle$ state.  For cross correlation measurements such as Fig.\ref{fig2}(a) an average of approximately 8000 experimental cycles were used.

The temperature of the cloud in the dipole trap is about 120 $\si{\micro}$K  corresponding to a measured atomic decoherence rate of $\gamma_0/2\pi\simeq100$ kHz, dominated by the Doppler broadening. The signal path detection efficiency is $q_s\simeq0.3$ including the fiber coupling efficiency and photodetector quantum efficiency.  The optical cavity has a waist size of $35~\si{\micro}$m, length of 13.7 mm, and out-coupling efficiency of 66\%. 

The single-photon Rabi frequency for $\sigma^+$ polarized light is $2g=2\pi\times 2.5$ MHz. Thus, the single-atom cooperativity for an atom on the cavity axis (along $z$) at an antinode of the cavity standing wave is given by $\eta=4g^2/\kappa\Gamma=8.6>1$, i.e. the system operates in the strong coupling regime of cavity quantum electrodynamics. The cavity resonance frequency matches the atomic frequency $|d\rangle\to|e\rangle$

{\bf {Detection and transmission probabilities.}}
The probability to observe a probe photon when a cavity photon is present and a signal photon is propagating through the EIT window at $\tau=0$ is given by \cite{Beck:prl2014}
\begin{eqnarray}
\varepsilon_0& = &\frac{1}{4}\frac{\eta^2}{(1+\eta)^2} \left[1-\exp\left(-\mathcal{D}/2\zeta\right)\right]^2,
\label{eq: epsilon}
\end{eqnarray}
where
\begin{eqnarray}
\zeta & = & \left(1+\frac{\gamma \Gamma}{\Omega^2}\right)\left(1+\frac{\Omega^2/\kappa\Gamma+\gamma/\kappa}{1+\eta}\right).
\end{eqnarray}
Here, $\eta=4.3$ is the spatially-averaged cavity cooperativity, $\mathcal{D}$ is the effective optical density that overlaps with the cavity mode, $\Omega$ is the control Rabi frequency, $\kappa= 2\pi\times140$ kHz is the decay rate of the cavity, $\gamma\simeq \gamma_c+\gamma_0$, $\gamma_0\approx2\pi\times100$ kHz is atomic decoherence rate in the absence of cavity photons, $\gamma_c$ is cavity-induced decoherence, and $\Gamma$ is the Cs excited-state decay rate. The decoherence rate, $\gamma_c$, caused by cavity light scattering manifests itself as: (1)~loss of atomic coherence given by $\langle n^{in}_c \rangle \kappa\eta/(1+\eta)^2$ where $\langle n^{in}_c \rangle$ is the mean $\sigma^+$-polarized input cavity photon number, (2)~reduction of signal transmission as a result of inhomogenous coupling of cavity light to atoms (see below). For the anti-correlation data shown in the inset of Fig.~2a, when we take into account the cavity blocking due to an atom in state $|d\rangle$, we obtain $4\varepsilon_0=0.1$ and a blocking probability for $\sigma^+$ light of $P=1-(1-\sqrt{4\varepsilon_0})^2=0.55$. This is in good agreement with the measured probability of $1-g^{(2)}_{s\sigma^+}(0)=0.59(7)$. A detailed theoretical treatment of the cavity interaction with atomic ensemble is given in Ref. [25]. 

In the nondestructive detection where horizontally-polarized cavity light is used, the detection probability is defined as the field amplitude of the transmitted $\sigma^+$ light, which interacts with atoms in state $|d\rangle$ as described in Ref. [23], combined with the field amplitude of $\sigma^-$ light on the output polarization beamsplitter. The field amplitude addition results in the factor 1/4 in Eq.~\ref{eq: epsilon}. In principle, this reduction can be avoided by impinging only $\sigma^+$ light onto an impedance-matched cavity and measuring the reflected photons. In our present lossy cavity, the reflection in the absence of signal photons causes a large background for the probe light.

Cavity-induced decoherence reduces the transmission probability of the signal photon and the EIT coherence time \cite{Beck:prl2014}. The signal transmission in the presence of cavity photons is given by:
\begin{eqnarray}
T_s = T_0\exp\left(-\frac{\mathcal{D}}{1+\Omega^2/\Gamma\gamma}\right)
\end{eqnarray}
where $T_0=\exp\left(-\frac{\mathcal{D}'}{1+\Omega^2/\Gamma\gamma_0}\right)$ is the EIT transmission corresponding to atoms outside the cavity waist and $\mathcal{D}'$ is the corresponding optical density.

An additional limit to the signal transmission is caused by the standing-wave nature of the cavity light in combination with the uniform distribution of atoms between nodes and antinodes of the cavity. Once the signal is detected, the spatial mode of the polation is projected onto the cavity mode resembling a grating imprinted onto the polariton structure. This effect leads to reduction in transmission of the signal. The overlap between the polariton before and after detection of a probe photon can be calculated as
\begin{eqnarray}
\mathcal{F}_p=\frac{1}{\pi}\int_0^{\pi}\frac{\eta\cos^2(\theta)}{1+\eta\cos^2(\theta)}d\theta=1- \frac{1}{\sqrt{1+\eta}}
\end{eqnarray}
where $\theta=kz$, $k$ is the wave-number of cavity light and $z$ is the position along the cavity axis. At large cooperativity, $\eta\gg1$, the expected maximum transmission approaches 100\%. For our system parameters this evaluates to about 70\%.

Also, the atomic cloud extended beyond the cavity region introduces additional signal transmission loss. This is because the signal photon wave-packet is localized inside the cavity region upon detection of a probe photon and therefore its spectral bandwidth exceeds the EIT bandwidth. Hence, after detection via the cavity, the signal photon propagating through the EIT window experiences dispersion and loss. Our numerical simulations predicts a loss of 30\% in signal transmission given the experimental parameters. In principle, this loss can be eliminated by removing atoms outside the cavity region. 

{\bf {Quantum correlation between probe and signal photons.}}
The mean photon rate entering the cavity can be calculated from the total detected photon rate exiting the cavity, $R^{s=0}_c$, in absence of signal photons as
\begin{eqnarray}
\langle R^{in}_c \rangle= \frac{R^{s=0}_c}{q_d(\frac{\mathcal{T}}{\mathcal{T}+\mathcal{L}})}
\end{eqnarray}
where $q_d=0.3$ accounts for detection losses including fiber coupling, filter losses and photodetector quantum efficiency and $\frac{\mathcal{T}}{\mathcal{T}+\mathcal{L}}=0.66$ is the cavity out-coupling efficiency with $\mathcal{L}$ and $\mathcal{T}$ being mirror loss and transmissivity, respectively. In the following, we combine $q_d$ and the cavity out-coupling efficiency into a single parameter $q_p$. The mean cavity photon number in absence of signal photons is then $\langle n^{in}_c \rangle =R^{in}_c \tau_c$ where $\tau_c=(\kappa/2)^{-1}$. The mean signal photon number in the relevant time window, i.e. the EIT life time $\tau_{_{EIT}}=(\Omega^2/(\Gamma\mathcal{D})+\gamma_0)^{-1}$, is given by $\langle n^{in}_s \rangle= R^{in}_s\tau_{_{EIT}}/q_s$ where $R^{in}_s/q_s$ is the signal photon rate entering the medium and $q_s=0.3$ accounts for detection losses. In absence of population in state $|d\rangle$, the linearly polarized cavity light is rotated by atoms in state $|g\rangle$ due to the differences in the coupling strengths for $\sigma^+$ and $\sigma^-$ polarized light interacting with state $|g\rangle$ and excited states. Ideally, this rotation is constant and we compensate for it with a waveplate at the output of the cavity. However, the shot-to-shot atom number fluctuation during loading provides a varying background, $\alpha q_p\langle n^{in}_c \rangle$, that dominates the probe port at low signal photon rates. We typically measure a maximum fractional background of $\alpha\approx3\times 10^{-3}$ of the total detected cavity photons. The detection events consists of a background given by $\langle b \rangle = \alpha q_p \langle n^{in}_c \rangle +\langle r_p \rangle$, where $\langle r_p \rangle$ denotes the dark counts of the probe detector $D_p$. We define the detected mean signal photon number $\langle n_s \rangle$, true detection events $\langle t \rangle$ and total detected mean probe photon number $ \langle n_p \rangle$ as
\begin{eqnarray} 
\langle n_s \rangle & = & q_s T_s\langle n^{in}_s \rangle + \langle r_s\rangle\\
\langle t \rangle & = & (\varepsilon_0 +\epsilon_b) q_p \langle n^{in}_c \rangle \langle n^{in}_s \rangle \\
 \langle n_p \rangle & = & \langle t \rangle + \langle b \rangle= (\varepsilon_0 \langle n^{in}_s \rangle+\epsilon_b \langle n^{in}_s \rangle+\alpha)q_p \langle n^{in}_c \rangle \nonumber\\
 &&+\langle r_p\rangle
 \end{eqnarray}
  where $\langle r_s\rangle$ denotes the dark-counts of the signal detector $D_s$ and $\epsilon_b=\varepsilon_d f_s$ is the probability of detecting a probe photon for a decohered atoms in state $|d\rangle$, $\varepsilon_d$, multiplied by the fraction of signal photons, $f_s$, incoherently mapped to state $|d\rangle$ via absorption. The coincidence counts are
  \begin{eqnarray}
  \langle n_sn_p \rangle & = & \varepsilon_0 q_p\langle n^{in}_c \rangle\times T_s q_s \langle n^{in}_s \rangle +\\ \nonumber 
 & & (\alpha+\epsilon_b \langle n^{in}_s \rangle) q_p\langle n^{in}_c \rangle\times T_s q_s \langle n^{in}_s \rangle +\nonumber \\
 && T_s q_s\langle n^{in}_s \rangle\langle r_p \rangle +((\varepsilon_0+\epsilon_b) \langle n^{in}_s \rangle+\nonumber \\
 && \alpha)\times q_p \langle n^{in}_c \rangle \langle r_s \rangle + \langle r_p \rangle \langle r_s \rangle. 
  \end{eqnarray}
Here, we assume that the conditional signal transmission is approximately equal to the mean signal transmission, $T_s$. Note that all terms, except the first, are caused by background sources. The cross-correlation function, neglecting the detectors' dark counts, can be approximated as
 \begin{eqnarray} 
  g^{(2)} (\tau=0)& = & \frac{ \langle n_sn_p \rangle }{ \langle n_s \rangle \langle n_p \rangle } \simeq \frac{1+\beta}{\beta+\langle n^{in}_s \rangle}
  \label{eq: g2}
\end{eqnarray}
where $\beta=\frac{\alpha+\epsilon_b \langle n^{in}_s \rangle}{\varepsilon_0}$.  When background processes are negligible ($\alpha, \epsilon_b, \langle r_s \rangle, \langle r_p \rangle \ll1 $), the maximum cross-correlation function at $\tau=0$ is simply approximated by $ g^{(2)} \simeq 1/\langle n^{in}_s\rangle$ for $\langle n^{in}_s\rangle<1$. Note that in the regime where $\langle r_{s} \rangle, \langle r_{p} \rangle\ll 1$, the correlation function $g^{(2)}$ is independent of the cavity photon number as both the detection probability and background scale linearly with it. However, the measured $g^{(2)}(\tau=0)$ drops at low cavity photon numbers where probe-part dark counts, $\langle r_p \rangle$, are not negligible compared to the detected cavity mean photon number. 

\begin{figure}[!t]
	\centerline{\includegraphics[width=\columnwidth]{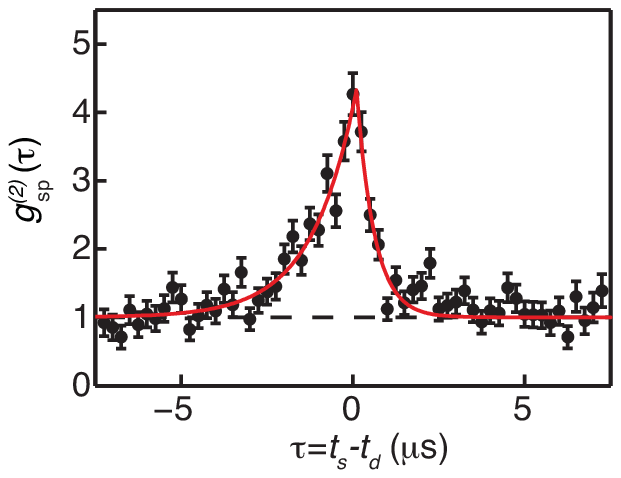}}
	\caption{Observed cross-correlation for double-pass signal beam, measured with $\langle n^{in}_c \rangle=4.4$ and $\Omega/2\pi= 2.9$ MHz. The fitted values are $g^{(2)}=4.4(5)$, $\tau_{<}=1.3(3)$ $\si{\micro}$s, and $\tau_{>}=0.5(2)$ $\si{\micro}$s. }
\label{fig: g2dp}
\end{figure}

To further increase the photon-photon interaction, we carried out an experiment to increase the effective optical density by transmitting the signal through the atomic ensemble twice. The retro-reflected signal is collected by a 90/10 fiber-beam splitter used at the signal input. We simultaneously measure auto-correlations of $g_{ss}^{(2)}=1.6(3)$, $g_{pp}^{(2)}=5.6(1)$ and the cross-correlation as plotted in Fig.~\ref{fig: g2dp}.\newline
\newline

\begin{figure}[!t]
\centerline{\includegraphics[width=\columnwidth]{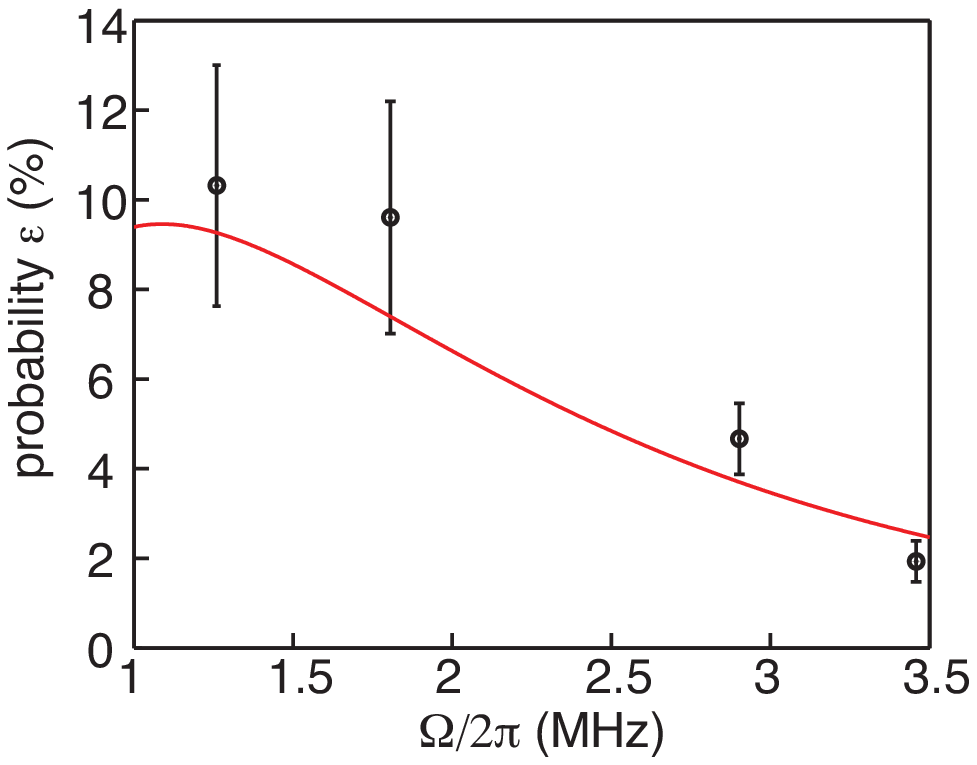}}
\caption{The total probability $\varepsilon$ calculated from the slope of the linear fits to the data in Fig.~3. The fitted curve represents the theory using Eq.\ref{eq: epsilon2} with fitted optical density $\mathcal{D}=4(2)$.}
\label{figs2: intg2}
\end{figure}

{\bf {Quantum efficiency.}}
The conditional nondestructive quantum efficiency of detecting a signal photon with mean input photon number $\langle n^{in}_s \rangle\ll1$ can be written as
\begin{eqnarray}
\text{Q} & = & \varepsilon q_p\langle n^{in}_c \rangle \simeq \frac{\langle n_sn_p \rangle- \langle n_p \rangle\langle n_s \rangle}{ \langle n_s \rangle}
\end{eqnarray}
where $\varepsilon$ is the total probability of having a probe photon given a signal photon traveling through the medium. It can be obtained from the asymptotic quantum efficiency and integrating the area under the $g^{(2)}$ function as 
\begin{eqnarray}
\varepsilon&=&\frac{\text{Q}}{q_p\langle n^{in}_c \rangle}= \frac{1}{q_p\langle n^{in}_c \rangle(1-\langle n^{in}_s \rangle)} \int{(g^{(2)}(\tau)-1) R_p d\tau} \nonumber \\
&=& \varepsilon_0\frac{\tau_c+\tau_{_{EIT}}}{\tau_c}.
\label{eq: epsilon2}
\end{eqnarray}
The probability $\varepsilon$ is calculated from the slope of the fitted lines in Fig.~4c and is plotted for different control Rabi frequencies in Fig.~\ref{figs2: intg2}. These extracted probabilities agree with theoretical predictions.

{\bf {Detection probabilities and QND requirements.}}
The QND requirements can be quantified using the measurement error, $\Delta X$, the transfer coefficient of input signal to meter (probe), $\mathcal{T}_M$, and transfer coefficient of input signal to output signal, $\mathcal{T}_S$ \cite{Grangier:Nature1998}. Using the formalism provided by Ralph {\it{et al.}} [Phys. Rev. A {\bf{73}}, 012113 (2006)], one can link the measurement probabilities in the discrete variable (DV) regime and $\mathcal{T}_S$ and $\mathcal{T}_M$ in continuous variable (CV) regime through different fidelity measures. The transfer coefficients in terms of measurement fidelity, $F_M$, and QND fidelity, $F_{QND}$, can be written as
\begin{eqnarray*}
\mathcal{T}_M &= & (\frac{2}{F_M^2} -1)^{-1}\\
\mathcal{T}_S &=&(\frac{2}{F_{QND}^2} - 1)^{-1}
\end{eqnarray*}
where 
\begin{eqnarray*}
F_M & = & P_{11}+P_{01} = \frac{\langle n_p\rangle}{\langle n^{in}_s\rangle}\\
F_{QND} & = & P_{11}+P_{10} = T_s.
\end{eqnarray*}
To estimate the measurement error in the CV regime, the conditional variance of the signal is measured and is compared to the shot-noise limit. In the DV regime, however, as the particle aspect of photons are detected and not the wave aspect, the conditional correlation function, $g^{(2)}_{ss|m}$ (signal auto-correlation function conditioned on detecting a meter photon), can be used instead to quantify the measurement error. In particular, a QND measurement satisfies $g^{(2)}_{ss|m}<1$ (quantum state preparation)and $\mathcal{T}_S+\mathcal{T}_M>1$. 

\end{document}